\documentclass{kapproc} 

\usepackage{procps} 

\usepackage[dvips]{graphicx}

\upperandlowercase

\let\footnote\savefootnote

\kluwerbib 

\newcommand\kms{{\rm{\,km\,s^{-1}}}}

\newcommand\Gyr{{\rm{\,Gyr}}}
\newcommand\apj{Ap.J.}
\newcommand\apjs{Ap.J.Supp}
\newcommand\apjl{Ap.J.Lett.}
\newcommand\aj{A.J.}
\newcommand\mnras{M.N.R.A.S.}
\newcommand\araa{Ann.Rev.Astr.\&Ap.}

\newcommand\apss{Ap.\&Sp.Sc.}
\newcommand\aap{A.\&Ap.}
\newcommand\aaps{A.\&Ap.Supp.}

\begin{document}

\articletitle{Extragalactic Thick Disks}

\articlesubtitle{Implications for Early Galaxy Evolution}

\author{Julianne J. Dalcanton (with Anil Seth \& Peter Yoachim)}
\affil{University of Washington, Seattle}
\email{jd@astro.washington.edu}

\begin{abstract}
  I briefly review the growing evidence that thick stellar disks
  surround most edge-on disk galaxies.  Recent studies show that these
  extragalactic thick disks have old ages, low metallicities, long
  scale lengths, and moderately flattened axial ratios, much like the
  thick disk of the Milky Way.  However, the properties of thick disks
  change systematically with the mass of the galaxy.  The thick disks
  of low mass galaxies are more prominent and somewhat more metal-poor
  than those surrounding massive disk galaxies.  Given the strong
  evidence that thick disks are fossils from an early epoch of
  merging, these trends place tight constraints on the early assembly
  of disk galaxies.
\end{abstract}


\section*{Introduction}

Although thick disks are best known as a component of the Milky Way,
they were actually first identified as extended faint envelopes
surrounding other galaxies (\cite{Burstein79,Tsikoudi79}).  The
subsequent discovery of a comparable stellar component in the Milky
Way (\cite{Gilmore83}) drew attention away from the ambiguous
interpretation of faint features in external galaxies, and instead
refocused it on the wealth of data available from photometric and
spectroscopic observations of individual stars within the Galaxy.
These studies quickly showed that the Milky Way's thick disk is a
fossil relic from an early epoch in the Galaxy's formation (see
reviews by \cite{Freeman02,Norris99,Majewski93,Gilmore89}).  Its stars
are old and metal-poor ($\langle$[Fe/H]$\rangle\sim$-0.7), and show an
enrichment pattern that is distinct from thin disk stars with similar
iron abundances (see the recent review by \cite{Feltzing04}).  These
data suggest that the Milky Way thick disk is not a simple extension
of the thin disk, but instead captures a unique episode early in the
formation of the Galaxy.

Given the evidence above, how does the thick disk fit into the existing
paradigm of disk galaxy formation?  The ages of stars in the thick
disk suggest that they were formed more than 8~Gyr ago (\cite{Liu00}),
when the merging rate was likely to be high.  It is therefore
reasonable to assume that the formation of the thick disk is somehow
coupled to the mergers and interactions expected to dominated in hierarchical
structure formation.  Within this scenario, thick disk stars could
have acquired their current large scale heights and vertical velocity
dispersions in three ways.  In the first and widely held view, the
thick disks stars were initially formed within a thin gas disk, but
were then vertically heated by one or more interactions with a
satellite galaxy (e.g.\ \cite{Quinn93}).  In the second, the thick
disk stars were formed {\emph{in situ}} at large scale heights during a
burst of star formation, as clumps
of gas coalesced to form the thin disk (e.g.\
\cite{Kroupa02,Brook04}).  In the third, the thick disk stars formed
outside of the galaxy in the pre-galactic fragments, which then 
deposited the stars in the disk (\cite{Statler88,Abadi03}).

Unfortunately, it is difficult to discriminate among these very
different scenarios using data from the Milky Way alone.  It is
therefore time to return to the broader range of galaxies that can be
probed with studies of extragalactic thick disks.  In the intervening
years between the initial detections of \cite{Burstein79} and
\cite{Tsikoudi79}, there has been a steadily growing body of
detections of thick disks in other galaxies (e.g.\
\cite{Neeser02,Wu02,Matthews00,Abe99,Fry99,Morrison97,Naeslund97,
deGrijs96,vanDokkum94,Bahcall85,Jensen82,vanderKruit81}; see also the
compilation in Table 2 of \cite{Yoachim05b} and the review by
\cite{Morrison99}).  Recently, the pace of discovery has accelerated
with the large early-type sample studied by \cite{Pohlen04} and the
late-type sample studied by \cite{Dalcanton02}.  In this review I
first summarize the structures, stellar populations, and kinematics of
the thick disk population, and argue that extragalactic thick disks
are indeed reasonable analogs of the well-studied thick disk in the
Milky Way.  I then discuss how the growing body of data places strong
constraints on the origins of thick disks, and on disk galaxy
formation in general.

\section{The Properties of Extragalactic Thick Disks}

\begin{figure}[ht]
\centerline{
\includegraphics[width=3.2in]{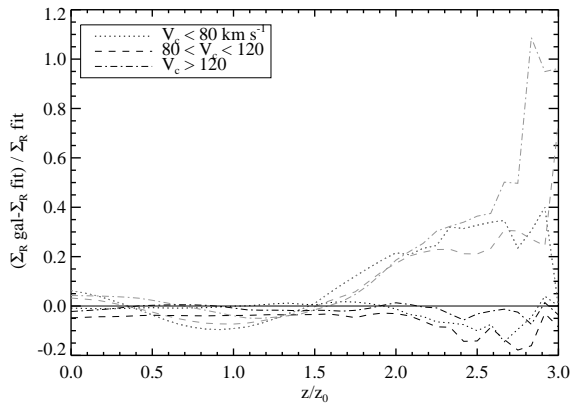}
}
\caption{Residuals from single-disk (light lines) and two-disk fits
  (dark lines) to edge-on galaxies in different mass ranges
  (\cite{Yoachim05b}).  All of the single-disk fits show large
  positive residuals at large scale-heights, for galaxies in every
  mass range.  In contrast, including a second thick disk component
  reduces the residuals to $<10$\% at all scale heights.}
  \label{residualfig}
\end{figure}

Until recently, most evidence for thick disks in other galaxies came
from deep broad-band imaging.  Starting with the seminal work of
\cite{Burstein79} and continuing through the systematic work of
Heather Morrison's group and others (see review in \cite{Morrison99}),
analyses of vertical surface brightness profiles (i.e.\ parallel to
the minor axis) of edge-on galaxies have typically shown breaks at
large scale heights, indicative of a second, thicker disk component
that dominates at large heights and faint surface brightnesses.  More
recent studies have taken advantage of advances in computing power to
fit the full two-dimensional light distribution to models of two
superimposed disk components (e.g.\ \cite{Yoachim05b,Pohlen04}).  In
Figure~\ref{residualfig} I show the residuals from one- and two-disk
fits to a large sample of edge-on late-type disk galaxies from the
\cite{Dalcanton00} sample, as analyzed in \cite{Yoachim05b}.  Fits to
only a single disk leave large amounts of light at high latitudes,
providing convincing evidence that there is an additional, thicker
stellar component.  Fits that include a second disk component do a far
better job of fitting the light distribution, at all latitudes.

\subsection{The Structure of Thick Disks}

Generically, all of the disk galaxies studied to date have
some degree of excess light at high latitudes.  While the
extra-planar light seems to be well fit by an additional thick disk
component, there is no guarantee that this second component is
strictly analogous to the well-studied thick disk of the Milky Way.
Within the Galaxy, one can identify thick disk stars by their distinct
kinematic and chemical properties, whereas two-dimensional
decompositions of galaxies typically have non-unique solutions that
depend on the weighting, the bandpass, and the assumed underlying
model.

In spite of these uncertainties, there is growing evidence that the
population of extragalactic thick disks revealed by two-dimensional
disk fitting are reasonable (though probably not exact) analogs of the
Milky Way's thick disk.  Structurally, the secondary disk components
are quite similar, and are well matched to the scale heights, scale
lengths, and axial ratios ($\sim$3-4:1) of the Milky Way thick disk,
when restricted to galaxies of comparable mass (\cite{Yoachim05b}).

\begin{figure}[ht]
\centerline{
\includegraphics[width=4in]{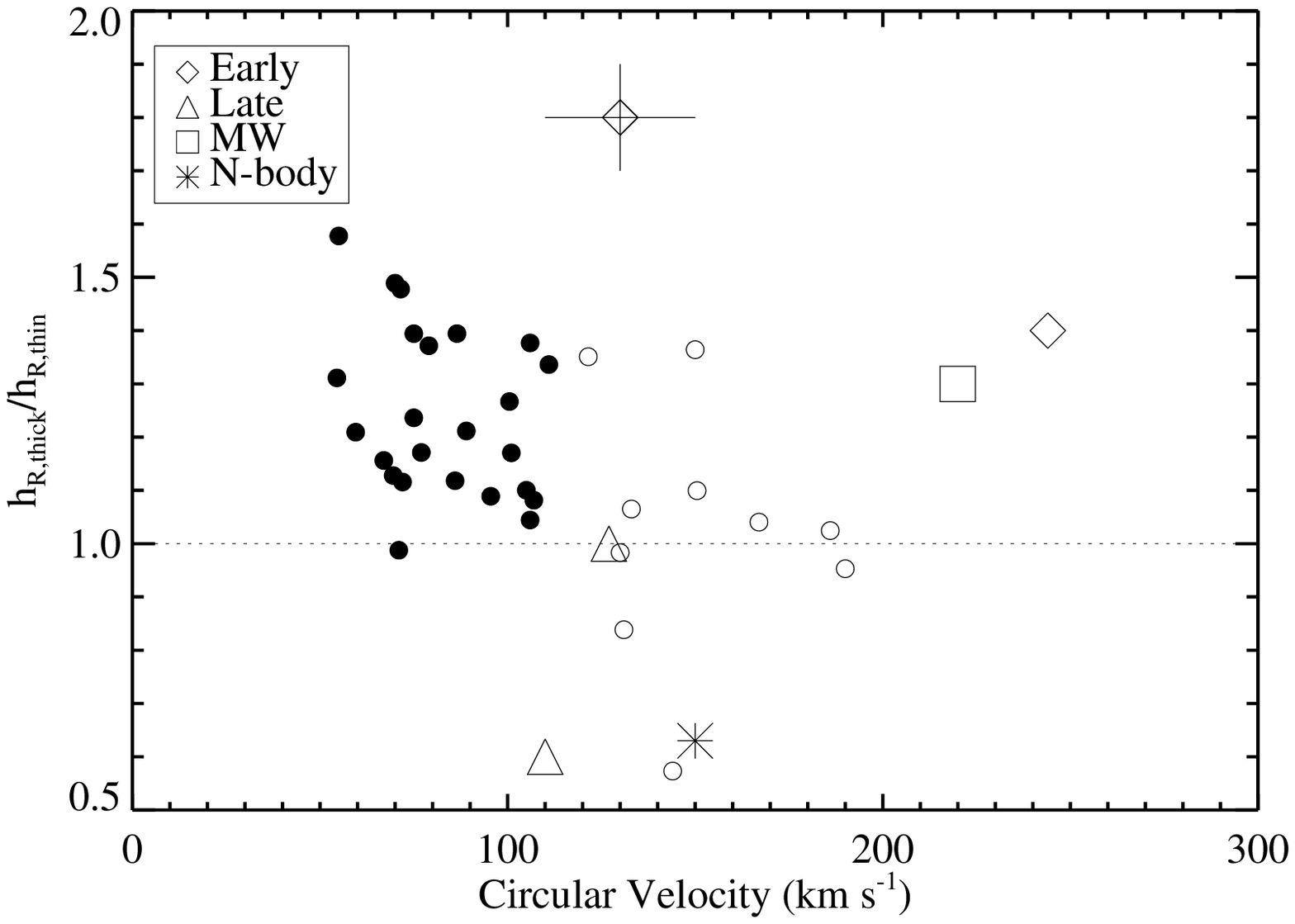}
} 
\caption{The ratio of thick disk to thin disk exponential scale
  length, as a function of galaxy rotation speed for the late-type
  galaxies from the \cite{Dalcanton00} sample (closed and open
  circles) and other individual studies (triangles;
  \cite{Wu02,Neeser02,Abe99}), the Milky Way (square; \cite{Larsen03}), the
  \cite{Pohlen04} sample of S0's, and numerical simulations of
  \cite{Brook04}. In almost all cases, the thick disks have longer
  scale lengths than their embedded thin disks.}\label{scalelengthfig}
\end{figure}

While it is not too surprising that extragalactic thick disks are
indeed thick, other unexpected structural results have been revealed
by the rapid increase in the number of well-studied systems.  In
particular, the thick disk component is almost always more
radially extended than the embedded thin disk.
Figure~\ref{scalelengthfig} shows the scale length ratio of the thick
to thin disks from a wide variety of studies in the literature.  In
almost all cases, the thick disk has a longer scale length.

\subsection{The Stellar Populations of Thick Disks}

\begin{figure}[ht]
\centerline{
\includegraphics[width=4in]{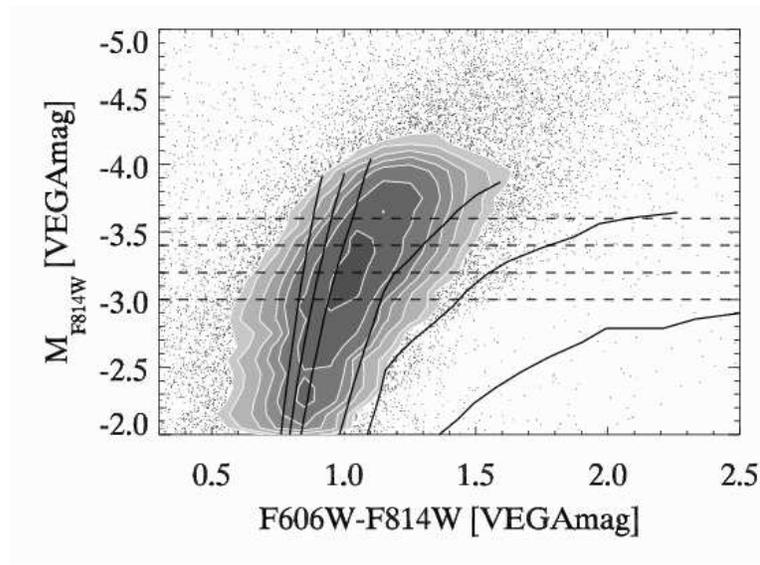}
}
\caption{The color-magnitude diagram of stars more than 4 disk scale
heights above the midplane, for all six galaxies in the \cite{Seth05b}
sample.  The stars are clearly dominated by old red-giant stars, with
no significant main sequence or AGB population.  Solid lines show 10
Gyr old Padova isochrones for the RGB with [Fe/H] (from left to right)
of -2.3, -1.7, -1.3, -0.7, -0.4, and 0.0.  The peak of the stellar
distributions fall between [Fe/H] of -1.3 and -0.7.  For reference,
the peak metallicities of the Milky Way's thick disk and stellar halo
are [Fe/H]$\sim$-0.7 and [Fe/H]$\sim$-2.2, respectively.}
\label{acsfig}
\end{figure}

In addition to their structural similarities, HST imaging is now
revealing similarities between the stellar populations of the Milky
Way and the extragalactic thick disks.  Studies of resolved
extraplanar stars (\cite{Seth05a,Seth05b,Mould05}) are finding that
the stellar populations are systematically older at large scale
heights.  Figure~\ref{acsfig} shows that stars above the plane are
dominated by old ($\sim5-13\Gyr$) red giant branch stars, as discussed
in more detail in \cite{Seth05b}.  The colors of these extraplanar red
giants indicate that they are exclusively metal poor, with a median
metallicity of [Fe/H]$\sim$-1 in the low mass ($V_c\sim75\kms$)
galaxies studied.  The old red giant population seems to be well mixed
vertically, and shows no evidence for the strong vertical metallicity
gradients that would be expected if steady vertical heating had been
solely responsible for driving each new generation of stars to larger
scale heights (\cite{Seth05b,Mould05}).  A single vertical heating
event, however, could be compatible with the lack of metallicity
gradient, since the resulting thickened disk would be well-mixed.

Unfortunately, extraplanar stellar populations have
been studied in fewer than ten galaxies to date, and usually in a
single field.  The number of future studies will be limited by the
relatively small numbers of galaxies that are both edge-on and close
enough to resolve into stars.

\subsection{The Kinematics of Thick Disks}

The number of kinematic studies of extragalactic thick disks is even
more sparse than the stellar population studies.  Outside the Milky
Way, there have only been two other published measurements of thick
disk kinematics (\cite{Yoachim05a}).  All three cases reveal thick
disks that rotate more slowly than the embedded thin disk.  However,
even within such a small sample, the thick disk kinematics seem to be
relatively complex.  Within the Milky Way, there is evidence at large
scale heights of significant kinematically distinct populations that
rotate even more slowly than typical thick disk stars closer to the
midplane (\cite{Gilmore02}).  Among the two extragalactic thick disks
studied, one shows clear evidence of {\emph{counter-}}rotation.  While
larger samples are clearly needed before one can probe systematic
variations, the current data already point to a significant degree of
scatter in the relative kinematics of thick and thick disks.

\section{How Do Thick Disks Form?}

The data summarized above points to the following generic facts about
thick disks that must be fit into the basic paradigm of disk galaxy
formation:

\begin{itemize}

\item Thick disks are found in essentially all edge-on disks, at all
  Hubble types.  They are therefore a generic by-product of disk galaxy
  formation.

\item Although they are ubiquitous, there is substantial scatter in
  the structural properties of thick disks (see, for example, the
  scatter in Figure~\ref{scalelengthfig}).

\item The thick disk typically has a longer exponential scale length
  than the embedded thin disk.

\item The stars in thick disks are dominated by old (>$5\Gyr$),
  relatively metal-poor (-1.5<[Fe/H]<-0.7) red giants, and are somewhat
  more metal-poor in lower-mass galaxies.

\item The kinematics of thick disk stars vary widely, and show both
  co- and counter-rotation.  However, in all cases the thick disk
  rotates more slowly than the thick disk.

\end{itemize}

Current models of disk formation assume that the bulk of a galaxy
disk forms from accreted gas with high angular momentum.  Analytic
calculations assume that this gas is accreted steadily through
spherical infall (e.g.\ \cite{Fall80,Dalcanton97,vandenBosch98}), but recent
numerical simulations have shown that even the gas accreted through
clumpy hierarchical merging can maintain sufficient angular momentum
to produce realistic massive disks (e.g.\ \cite{Governato04}).  The
accreted gas dissipates and collapses into a rapidly rotating disk that
then converts into stars, forming a thin disk.  

Within this scenario, the natural sites of thick disk star formation
are: [1] in the thin gas disk, which is later disrupted and vertically
heated by satellite accretion; [2] in the merging gas clumps, as they
coalesce into the thin disk (\cite{Brook04}); or [3] in the
pre-galactic fragments, before the disk is formed (e.g.\
\cite{Abadi03,Yoachim05b}).  All three of these possibilities would
naturally lead to the formation of the thick disk in a merging
hierarchy.  Moreover, because all three are tied to high merging rates
(which peak early and decline sharply after a redshift of $\sim1-2$),
each of these scenarios naturally produces an early epoch of rapid
thick disk formation, making them compatible with the old ages and
high $\alpha$-element abundances currently seen in thick disk stars.

Of these three scenarios, I believe that the third possibility --
direct accretion of stars -- is likely to dominate the production of
the thick disk, although there is certainly room for the other two
mechanisms to occur as well (and indeed, it would be surprising if
they didn't to some degree).  Although it is far more limited than the
structural information currently in hand, the kinematic data strongly
disfavors the vertical heating scenario.  Vertical heating does little
to change the angular momentum of a disk (\cite{Velazquez99}), thus
making it difficult to explain why the thick disk rotates
significantly more slowly than the thin disk, having no more than
$\sim$50\% of the rotational speed of the thin disk in all three cases
studied.  Likewise, it would be difficult to produce the very
different scale lengths of the two disk components without their
having very different angular momenta.  

Long thick disk scale lengths are likewise difficult to produce in the
second model, where thick disk stars form {\emph{in situ}} as gas
merges to form the final disk.  In this case, the same accreted gas
would form both the thin disk and the thick disk stars, making it
likely that both would have similar radial distributions.  Producing a
thin disk with a shorter scale length would require that the gas left over
after the thick disk stars formed had systematically lower
angular momentum.  This requirement seems a bit unlikely and is in
conflict with the limited available kinematic data.

In contrast, the slower rotation speeds and longer scale lengths of
thick disk stars drop out naturally if the stars were formed in
pre-galactic fragments before being accreted onto the disk.  Gas and
stars behave differently during accretion, leading to a natural
segregation in the properties of the thin and thick disk.  The
accreted gas (which forms the thin disk) can cool and dissipate, and
thus will tend to contract further into the halo than the stars.  As
it contracts, the gas will speed up due to angular momentum
conservation, producing a compact disk that rotates more rapidly than
the accreted stars.  In contrast, the orbits of the accreted stars
cannot lose energy (except through dynamical friction), and will tend
to remain in a thicker, more radially extended distribution than the
gas disk.  This scenario thus naturally reproduces the larger scale
height and scale length of the thick disk.  Moreover, because not all
of the merging satellites have the same gas-richness, some satellites
may contribute the bulk of the stars to the thick disk while others
contribute the bulk of the gas to the thin disk.  Some kinematic and
structural decoupling between the two components would therefore be
expected, because different precursors may dominate production of the
thick and the thin disks, and could thus easily produce the scatter
seen in the observations.

\begin{figure}[ht]
\centerline{
\includegraphics[width=4in]{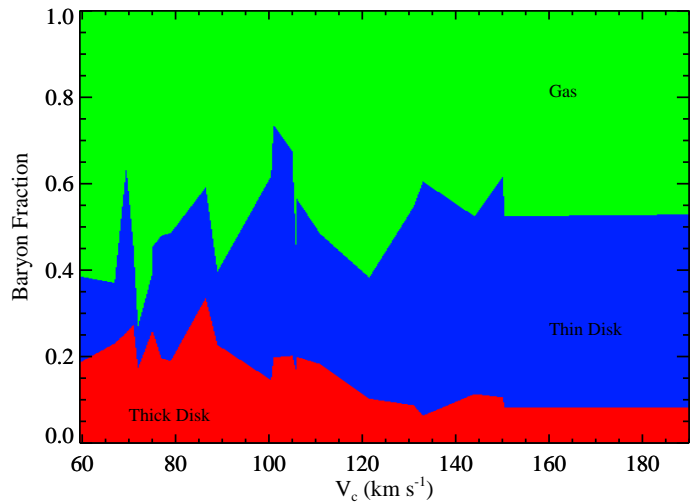}
} 
\caption{The baryonic mass fraction of the the thick disk stars
  (bottom), thin disk stars (middle), and cool gas (top) for the
  galaxies analyzed by \cite{Yoachim05b}, sorted by galaxy rotation
  speed.} \label{baryonfig}
\end{figure}

If the majority of thick disk stars were indeed directly accreted,
then the thick disk can constrain the typical gas-richness of the
pre-galactic fragments from which the galaxy assembled.  In this
picture, any stars in the merging sub-units were deposited in the
thick disk, while any gas cooled and settled into the thin disk.  The
relative baryonic mass fractions of the thin and thick disks therefore
constrain the ratio of stars to gas in the early galaxy.  In
\cite{Yoachim05b} we have inferred the luminosities of the thick and
thin disks from the two-dimensional decompositions, and then adopted
color-dependent stellar mass-to-light ratios to derive the stellar
mass of each disk component.  The resulting ``baryon budget'' is
reproduced in Figure~\ref{baryonfig}, as a function of galaxy mass for
the late-type \cite{Dalcanton00} sample, after assuming that any gas
is confined to the thin disk.

Figure~\ref{baryonfig} shows a number of striking features.  First,
the baryonic mass fraction trapped in the thick disk is relatively
small, implying that the initial galaxy disk was almost entirely
gaseous ($\sim$75-90\%, or more if some thick disk stars formed during
the final coalescence of the disk).  Second, the stellar mass of the
thick disk is {\emph{larger}} than the stellar mass of the thin disk
in low mass galaxies.  Thus, although low mass galaxies appear to be
blue and young, the majority of their stars are actually quite old, as
would be expected in hierarchical merging scenarios.  Finally, the
thick disk is increasingly important in low mass galaxies.  The
baryonic mass fraction locked in the thick disk increases
systematically from $\sim10$\% in high mass galaxies to up to
$\sim25$\% in low mass galaxies, implying that the precursors of low
mass galaxies were more gas poor than the precursors of their high
mass counterparts.

If thick disk stars are directly accreted, then the increasing
importance of thick disk stars in low mass galaxies can be easily
explained by supernova-driven winds in the pre-galactic fragments.
The precursors of low mass galaxies are themselves likely to be lower
mass than the typical sub-units which merge together to form more
massive galaxies, and thus, they are more susceptible to gas loss
through supernova driven winds.  These winds would reduce the gas mass
of the initial galaxy, and thus reduce the fraction of baryons that
wind up in the thin disk.  The winds would also reduce the typical
metallicity of thick disk stars, and could thus nicely produce the
trend toward lower thick disk metallicities in low mass galaxies
(\cite{Seth05b}).

\section{Prospects for the Future}

Decades of painstaking work have gone into observing and characterizing
the population of extragalactic thick disks.  Thanks to this effort,
we now have a broad understanding of the basic properties of thick
disks, allowing us to finally place them within the context of galaxy
formation.  However, while the broad outlines of thick disk formation
are in place, there are many details to be worked out.  Over the next
decade, I believe there are several areas where thick disk
research will be critical:

\begin{itemize} 

\item{\bf{Exploring the link between observations of thick disks at
      low redshift and of young galaxies at high redshift ($z>1.5$).}}
  There are already hints that many of the stars in clumpy, disturbed
  galaxies at high redshift will evolve into the thick disk population
  by the present day (\cite{Elmegreen05}).  This likely connection
  makes on-going work on low-redshift thick disks crucial for
  understanding and interpreting high-redshift observations.

\item {\bf{Linking thick disk observations to numerical studies of
      galaxy formation.}}  Simulations are just now beginning to have
  the resolution to tackle thick disk formation (see work by Chris
  Brook and Alyson Brooks in this volume).  However, because the
  properties of thick disks depend on star formation and
  gas loss in dense low mass pre-galactic fragments, simulations of
  thick disk formation will be highly sensitive to the adopted star
  formation and feedback recipes, as well as to the numerical
  resolution.  Thus, I believe we are only in the early days of
  numerical exploration of thick disk formation.  Perhaps the greatest
  long-term role for extragalactic thick disks will be as a key
  calibration for tuning the baryonic physics in these simulations.

\item {\bf{Incorporating the thick disk component into semi-analytic
      models of galaxy formation.}}  Given that the thick disk
  dominates the stellar mass in low-mass disks, a recipe for including
  thick disks in semi-analytic models is sorely needed.

\item {\bf{Exploring the connection between thick disks and bulges.}}
  To date, large studies of thick disks have focused on either
  bulgeless (\cite{Dalcanton02,Yoachim05b}) or bulge-dominated
  (\cite{Pohlen04}) galaxies.  However, there is currently no study
  using uniform data and analysis techniques to bridge across this
  wide range of galaxy types.  There are hints that thick disks may be
  systematically ``thicker'' in early type galaxies
  (\cite{Yoachim05b}), but without uniform data, the robustness and
  interpretation of such conclusions is questionable.  Theoretically,
  there is much work to be done as well.  Mergers are thought to be
  critical to the formation of both thick disks and massive bulges,
  but we currently have no well-developed theory that can reliably
  calculate how the merging material is distributed between the bulge
  and the thick and thin disks.

\item {\bf{Observationally and theoretically untangling thick disks
      and stellar halos.}}  As difficult as it has been to
  characterize extragalactic thick disks, the upcoming work to isolate
  stellar halos will be even harder.  In the Milky Way, the thick disk
  makes up roughly 10\% of the stars at the solar circle, while the
  stellar halo contributes only a tenth of a percent (\cite{Chen01}).
  Even with detailed studies of resolved stars (e.g.\
  \cite{Ferguson02}), it is difficult to decide whether a given
  structure is an analog of the Milky Way's thick disk or its halo, if
  indeed they are truly distinct structures.  Already, there have been
  extraplanar stars attributed to the thick disk that are probably
  distributed in a much more spherical distribution (e.g.\
  \cite{Tikhonov05}) and which may be better assigned to an analog of
  the stellar halo.  Simultaneously, theoretical models of stellar
  halo formation (\cite{Bullock05}) produce distributions of stars
  that can be relatively flattened in their inner, higher surface
  brightness regions.  Given the only modest flattening ($\sim$3:1)
  of extragalactic thick disks, it is possible that
  these are the same structures, in spite of the different
  nomenclature.  As more simulations are translated to the
  observational plane, the correct interpretation of broad-band colors
  and resolved stellar populations in edge-on disks will clarify.

\end{itemize}

\end{document}